# An Overview of 3GPP Release-15 Study on Enhanced LTE Support for Connected Drones

Siva D. Muruganathan†, Xingqin Lin†, Helka-Liina Määttänen†, Jonas Sedin†, Zhenhua Zou†, Wuri A. Hapsari‡, Shinpei Yasukawa‡

†Ericsson, ‡NTT DOCOMO

Contact: xingqin.lin@ericsson.com

***Abstract*—Cellular connectivity to low altitude unmanned aerial vehicles (UAVs) has received significant interest recently which has led to a 3GPP study on enhanced LTE support for connected UAVs in Release 15. The objective of the study is to investigate the capability of long-term evolution (LTE) networks for providing connectivity to UAVs. In this article, we provide an overview of the key findings of the 3GPP Release-15 study-item phase. We first introduce UAV connectivity requirements and performance evaluation scenarios defined in the study. We then discuss radio channel models and the key identified challenges of using LTE networks to provide connectivity to UAVs. We summarize potential solutions to address the challenges including interference detection and mitigation techniques, mobility enhancements, and UAV identification. Finally, we also shed light on the key features standardized during the Release-15 work-item phase.**

## I. Introduction

Most of the activities in terms of research and development in the area of mobile broadband have been focused on providing wireless broadband communication to outdoor users on the ground or indoor users in buildings. Recently, providing cellular connectivity to low altitude unmanned aerial vehicles (UAVs, a.k.a. drones) has gathered increasing interest from the industry [1]-[4], academia [5]-[6], and public policy makers [7]-[8]. There is a plethora of use cases for supporting aerial vehicles in cellular networks, some of which include search-and-rescue, surveillance, wildlife conservation, package delivery, and monitoring of critical infrastructure [9].

When aerial vehicles are flying well above eNodeB antennas, they may have a high likelihood of line-of-sight (LOS) propagation conditions to multiple neighbouring eNodeBs. In such a scenario, an uplink signal transmitted from an aerial vehicle may become visible and cause interference to multiple neighbouring eNodeBs [10]. If this interference is not controlled or mitigated, it may adversely impact the uplink performance of existing users on the ground. Hence, to protect the existing users in the LTE network, the network may have to perform certain actions such as ensuring that the interference is mitigated or performing admission control of aerial vehicles in the network. As a prerequisite to either of these actions, the network may first need to identify the aerial vehicles. With the abovementioned LOS propagation conditions, downlink signals transmitted from multiple eNodeBs may cause downlink interference to the aerial vehicle. Another issue is whether the existing mobility mechanism of LTE networks is sufficient or whether it needs enhancements to support cellular connectivity of aerial vehicles.

To better answer these issues and to understand the potential of LTE networks for providing cellular connectivity to aerial vehicles, the 3$^{rd}$ generation partnership project (3GPP) started the Release-15 study on enhanced LTE support for aerial vehicles in March 2017 [11]. This study assessed the performance of LTE networks supporting aerial vehicles with up to Release-14 functionality. The study was completed in December 2017 and the outcomes are documented in the 3GPP technical report TR 36.777 [1] including comprehensive analysis, evaluation, and field measurement results. With the completion of this study-item, 3GPP started a follow-up work-item [12] phase to specify certain features to provide more efficient cellular connectivity to aerial vehicles. The work-item phase for enhanced LTE support for connected aerial vehicles was concluded in June 2018.

In this article, we provide an overview of the 3GPP Release-15 study on enhanced LTE support for aerial vehicles and shed light on the key features standardized during the Release-15 work-item phase. The overview provided by this article is an accessible first reference for researchers interested in learning the 3GPP state-of-the-art findings on cellular connected drones.

## II. Requirements, Scenarios, and Channel Models

### A. Performance Requirements and Deployment Scenarios

The two main data types with regards to wireless connectivity of aerial user equipments (UEs) are command-and-control data and application data. The ability to send command-and-control traffic to aerial UEs from eNodeBs can significantly improve the safety and operation of aerial UEs. For instance, it is critical that information such as changes in the flight route are conveyed to the aerial UEs in a timely manner with sufficient reliability. In the 3GPP Release-15 study, uplink/downlink data rate requirement of 60-100 Kbps was defined for command-and-control traffic to ensure proper operational control of aerial UEs. In addition, the Release-15 study also defined up to 50 Mbps as uplink data rate requirement for application data.

To evaluate the performance of LTE networks in the presence of LTE connected aerial vehicles, the following three scenarios were considered:

- Urban-macro with aerial vehicles (UMa-AV)
- Urban-micro with aerial vehicles (UMi-AV)
- Rural-macro with aerial vehicles (RMa-AV)

UMa-AV represents scenarios where the eNodeB antennas are mounted above the rooftop levels of surrounding buildings in urban environments. Urban scenarios with below rooftop eNodeB antenna mountings are represented by UMi-AV. Larger cells in rural environment with eNodeB antennas mounted on top of towers are represented by RMa-AV.

In the evaluations during the Release-15 study, aerial vehicles were modeled as outdoor UEs with heights well above ground level, while terrestrial users were on the ground and inside buildings. The total number of UEs (including both aerial UEs and terrestrial UEs) per cell was assumed to be 15. To study the impact of supporting aerial UEs with different densities in a cell, aerial UE ratios of 0%, 0.67%, 7.1%, 25%, and 50% were considered.

### *B. Channel Modelling*

To characterize the channels between aerial UEs and eNodeBs, the Release-15 study defined models for LOS probability, pathloss, shadow-fading, and fast-fading. In this section, we highlight some key aspects of these models.

To define LOS probability, an aerial UE height dependent modelling approach was adopted in the study. For all three scenarios, the LOS probability models defined in [13] were reused for aerial UE heights below a lower height threshold. The lower height threshold is 22.5 m for UMa-AV and UMi-AV, while the lower height threshold is 10 m for RMa-AV.

Since eNodeB antennas are well above rooftop in UMa-AV and RMa-AV, a 100% LOS probability is assumed above an upper height threshold. The upper height thresholds are 100 m and 40 m for UMa-AV and RMa-AV, respectively. As eNodeB antennas are below rooftop in UMi-AV, the probability of non-line-of-sight (NLOS) is generally higher in UMi-AV when compared to UMa-AV and RMa-AV. Hence, the upper height threshold is not applicable in UMi-AV.

In the aerial UE height range between the lower and upper height thresholds, the LOS probability models in [1] were derived via ray tracing simulations for UMa-AV and RMa-AV. For UMi-AV, ray tracing simulations were used to derive a LOS probability model that applies for aerial UE heights above the lower height threshold.

For all three scenarios, the pathloss and shadow-fading models defined in [13] are reused for aerial UE heights below a lower height threshold. The lower height threshold is 22.5 m for UMa-AV and UMi-AV, while the lower height threshold is 10 m for RMa-AV. In the aerial UE height range above the lower height threshold, pathloss and shadow-fading models for both LOS and NLOS conditions were agreed in the Release-15 study considering field measurements and ray tracing simulation results contributed by multiple sources.

Three alternative fast-fading models were agreed during the Release-15 study. The three alternatives differ in the angular spreads, delay spreads, and K-factor ranges as well as modelling methodology.

## III. Problems Identified During The Study Phase

During the Release-15 study, interference problems were identified in both uplink and downlink for scenarios involving aerial UEs. In this section, we highlight the identified uplink and downlink interference problems.

### *A. Uplink Interference*

In the uplink, the aerial UEs were found to cause interference to more cells than a typical terrestrial UE could. This is because aerial UEs, when they are airborne, experience LOS propagation conditions to more cells with higher probability than terrestrial UEs. This generally translates into higher interference caused by an airborne aerial UE to these cells. The uplink interference over thermal-noise (IoT) ratios for terrestrial UEs are given in Figure 1(a) which shows the effect of increased uplink interference on terrestrial UEs as the aerial UE ratio increases.

Due to the increased uplink interference, the uplink throughput performance of terrestrial UEs degrades when the aerial UE ratio is increased in the network. The degraded uplink throughput of terrestrial UEs in turn increases the uplink resource utilization level in the network. In other words, degraded uplink throughputs imply that the UEs take longer time to transmit their data which will consume more resources and will lead to increased uplink resource utilization. An increased uplink resource utilization level inherently means an increased level of uplink interference in the network, which in turn was observed to degrade the uplink performance of both aerial UEs and terrestrial UEs.

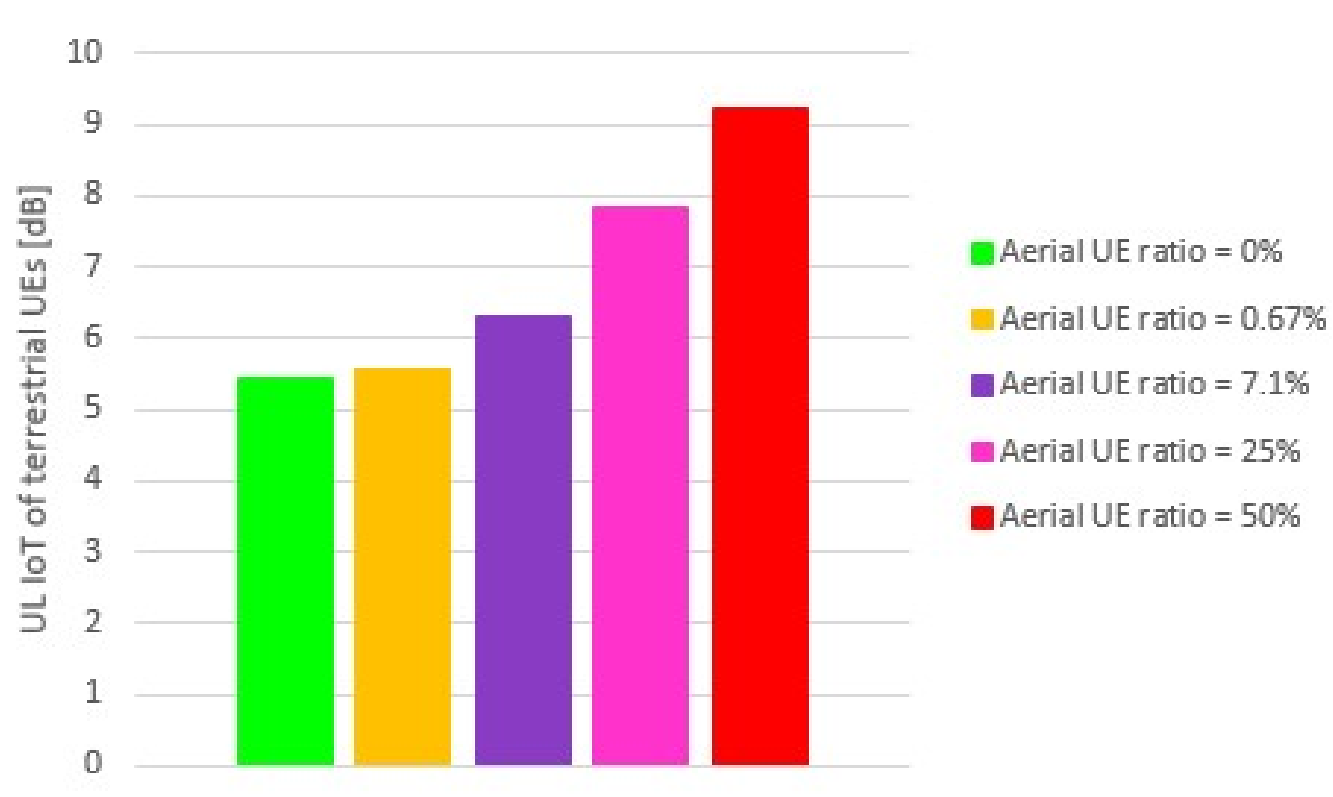

(a)

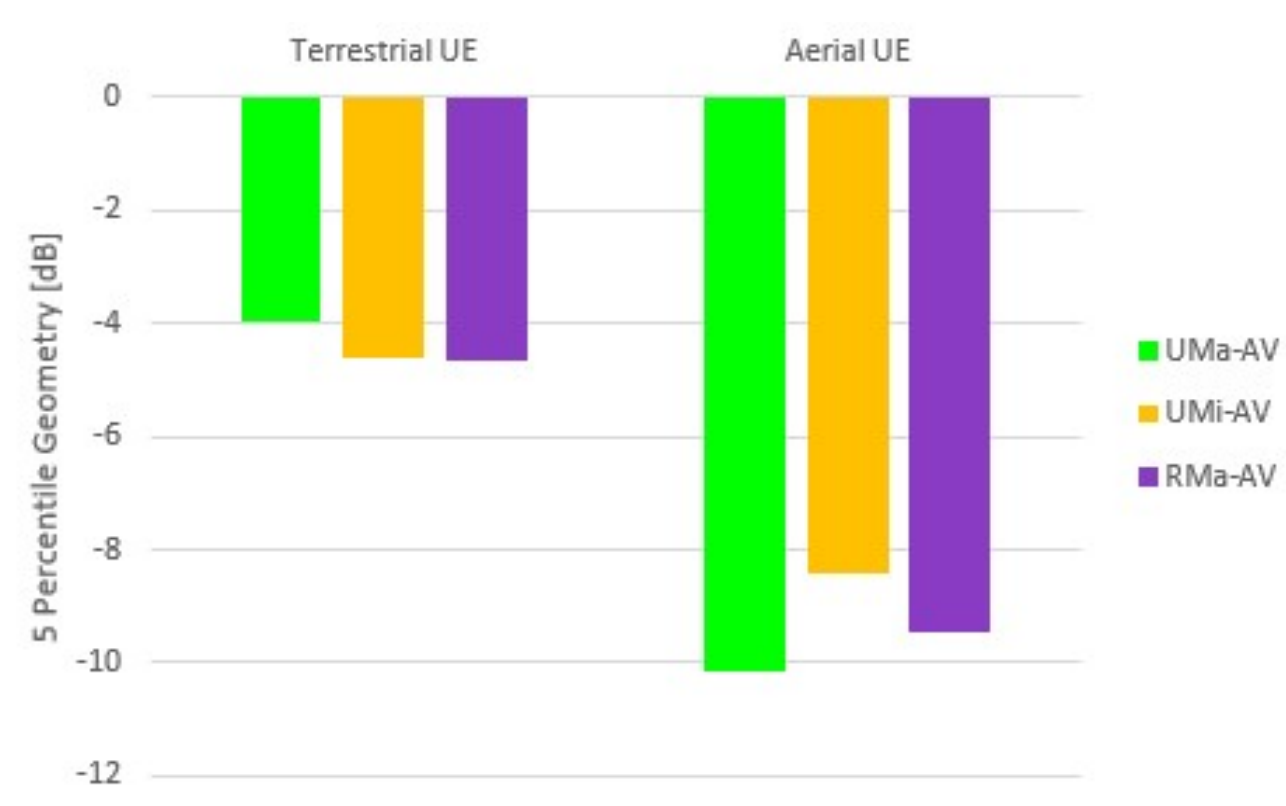

(b)

Figure 1. Illustration of UL and DL interference: (a) increase in UL IoT of terrestrial UEs with increasing aerial UE ratios; (b) comparison of five-percentile geometry showing poorer geometries of aerial UEs when compared to those of terrestrial UEs.

### B. Downlink Interference

In the downlink, compared to a typical terrestrial UE, the aerial UEs observe interference from more cells due to the LOS propagation conditions experienced by aerial UEs when they are airborne. Figure 1(b) compares the five-percentile downlink geometry experienced by the aerial UEs when compared to that experienced by terrestrial UEs. Here geometry is defined as the ratio of the average received power from the serving cell to the sum of the average interference power and noise power. The average interference power is computed as the sum of average received power from all non-serving cells. The degraded downlink geometry experienced by the aerial UEs is a result of receiving downlink inter-cell interference from multiple cells.

Due to the increased downlink interference, the downlink throughput performance of aerial UEs degrades. The degraded downlink throughput of aerial UEs when coupled with increased aerial UE ratios increases the downlink resource utilization level in the network. An increased downlink resource utilization level inherently means an increased level of downlink interference in the network, which in turn degrades the downlink performance of both aerial UEs and terrestrial UEs.

To address the challenges due to the uplink and downlink interference, 3GPP studied interference detection and mitigation techniques, mobility enhancements, and UAV identification. Some solutions were standardized to address these challenges during the Release-15 work-item phase, while some issues were concluded to be solvable via implementation-based techniques. In the next sections, we provide an overview and shed light on the solutions standardized during the Release-15 work-item phase.

## IV. Solutions for Interference Detection

Interference detection is a useful prerequisite for applying interference mitigation. In this section, we first discuss the potential solutions considered during the study-item phase and then describe the features standardized in the work-item phase for interference detection.

### A. UE-based Solutions Considered During Study-Item Phase

In an LTE network, a UE can be configured to perform neighbouring cell measurements such as reference signal received power (RSRP), reference signal received quality (RSRQ), and reference signal–signal to interference plus noise ratio (RS-SINR). Downlink interference can be detected based on these UE measurements reported to the eNodeB. One key aspect is to link the triggering of measurement reports to the changing interference conditions. An enhanced triggering condition may be a function of more than single cell RSRP. For example, the measurement report can be triggered when multiple cells' RSRP/RSRQ values are above/within a threshold or when a sum of multiple cells' RSRP/RSRQ values is above a threshold. Uplink interference can be detected either based on measurements at eNodeB or based on measurements reported by the UE.

Potential enhancements that were considered included new triggering events as discussed above, enhancements of triggering conditions, and the inclusion of more results in the measurement report. In addition, UE related information such as mobility history reports, speed estimation, timing advance adjustment values, and location information might also be useful for interference detection.

### B. Network-based Solutions Considered During Study-Item Phase

For network-based solutions, one candidate relied on information exchange between the eNodeBs. For example, measurements reported by UE are exchanged between eNodeBs when feasible and are used for interference detection. Another example is exchanging uplink reference signal configuration information of aerial UEs. By exchanging this information, a neighbouring eNodeB can measure the uplink interference caused by an aerial UE via measuring the power of the uplink reference signal.

Information on eNodeB's downlink transmission power can also be exchanged between eNodeBs. With neighbor eNodeB's transmission power, the uplink pathloss between an aerial UE and the specific target neighbor eNodeB can be determined from the UE's measurement reports assuming reciprocity. The uplink interference then can be estimated from the transmission power of the UE and the uplink pathloss.

It should be noted, however, that the feasibility of exchanging all this information depends on factors such as the type of backhaul and the feasibility of exchanging such information over a large number of eNodeBs.

### C. Interference Detection Features Specified during Work-Item Phase

One of the features specified during the Release-15 work-item phase involves a UE sending a measurement report to the network when the UE sees a certain event criterion being fulfilled simultaneously for multiple cells. In other words, the UE is configured with a triggering condition such that the UE reports a measurement when a certain number of cells $X$ fulfill an entry condition of a given event during a time to trigger (TTT[1]). This feature is useful for detecting interference from a flying aerial UE. Upon reception of such a measurement report, the network can detect interference early and apply different interference mitigation solutions. For the UE to report measurements when event criterion is met for multiple cells, the following need to be configured to the UE: (a) number of cells $X$, (b) the event to use which can be from events A3, A4, and A5 as defined in [15][2], and (c) the thresholds associated with events.

An illustration of UE measurement reporting when an event is triggered for multiple cells is shown in Figure 2. In this illustration, the number of cells $X$ for which an event must be fulfilled before triggering a measurement report is set to two. As shown in the figure, when preconfigured events are fulfilled for cells A and B, these cells are added to the 'cellsTriggeredList'. After both cells A and B are added to the 'cellsTriggeredList' (i.e., corresponding to *X=2* cells), a measurement report is triggered from the UE to the network which can be used for interference detection. It should be noted that when preconfigured events are subsequently fulfilled for more than *X=2* cells (for example, in a third cell which is cell D in Figure 2), a measurement report will not be triggered by the UE.

[1] According to the 3GPP specification [15], TTT is defined as 'the time during which specific criteria for the event need to be met in order to trigger a measurement report.'

[2] According to the 3GPP specification [15], event A3 is defined as the event when a neighbor cell becomes a configured offset better than the serving cell. Event A4 is the event that a neighbor cell becomes better than an absolute threshold. Event A5 is the event where the serving cell becomes worse than one absolute threshold and the neighbor cell becomes better than another absolute threshold.

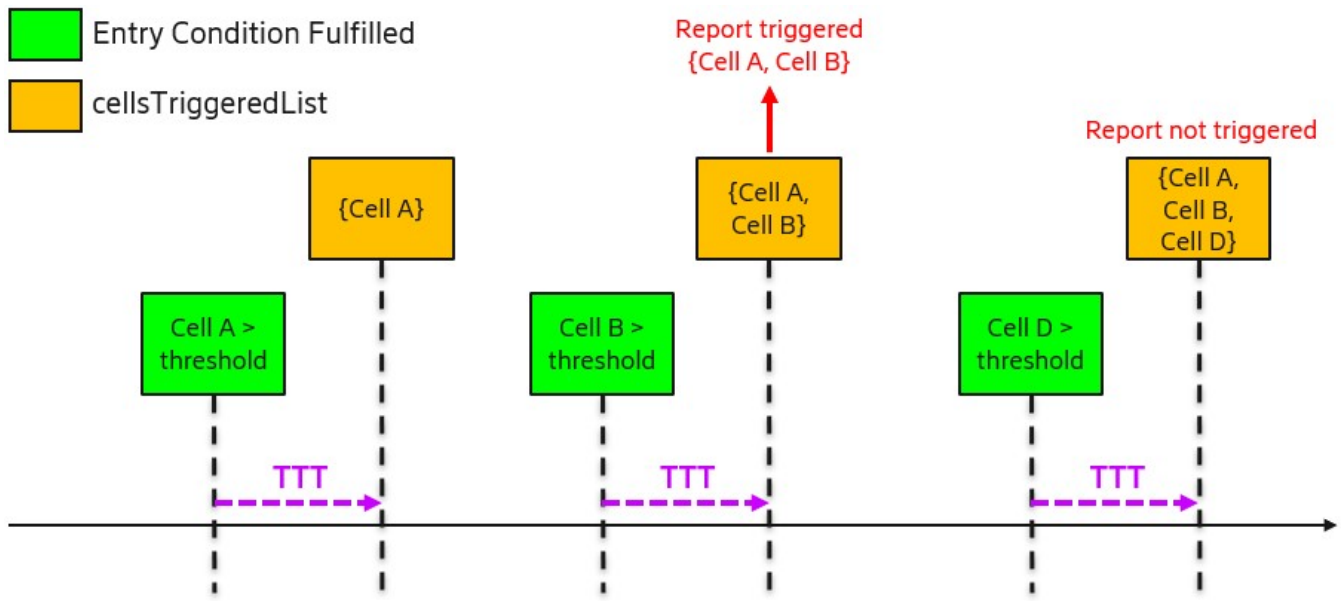

Figure 2. An illustration UE measurement reporting when a certain preconfigured event is triggered for multiple cells

Figure 3 shows a numerical example where the UE is configured to report a measurement when 4 cells fulfill entry conditions for event A4. In this figure, RSRP values corresponding to a set of different cells over time for an aerial UE moving at a speed of 150 km/h with a 200 m altitude are shown. The absolute threshold associated with the A4 event is -76 dBm in the example. As shown in Figure 3, RSRP measurements corresponding to four neighbouring cells become better than the threshold of -76 dB near 4.4 s. Hence, the aerial UE will trigger a measurement report shortly after 4.4 s to the network which can be used for interference detection.

A second feature specified during the Release-15 work-item phase is height-based measurement reporting when an aerial UE crosses a height threshold. A height-based measurement report may include the height, the location, and the horizontal/vertical speeds of the aerial UE. The height threshold is configured to the aerial UE via higher layer signaling. The height-based reporting can be useful for interference detection. When an aerial UE is flying above a certain height, both the uplink and downlink interference increase. By configuring an aerial UE with appropriate height threshold, the network can detect interference early when an aerial UE crosses the configured height threshold and sends a height-based report.

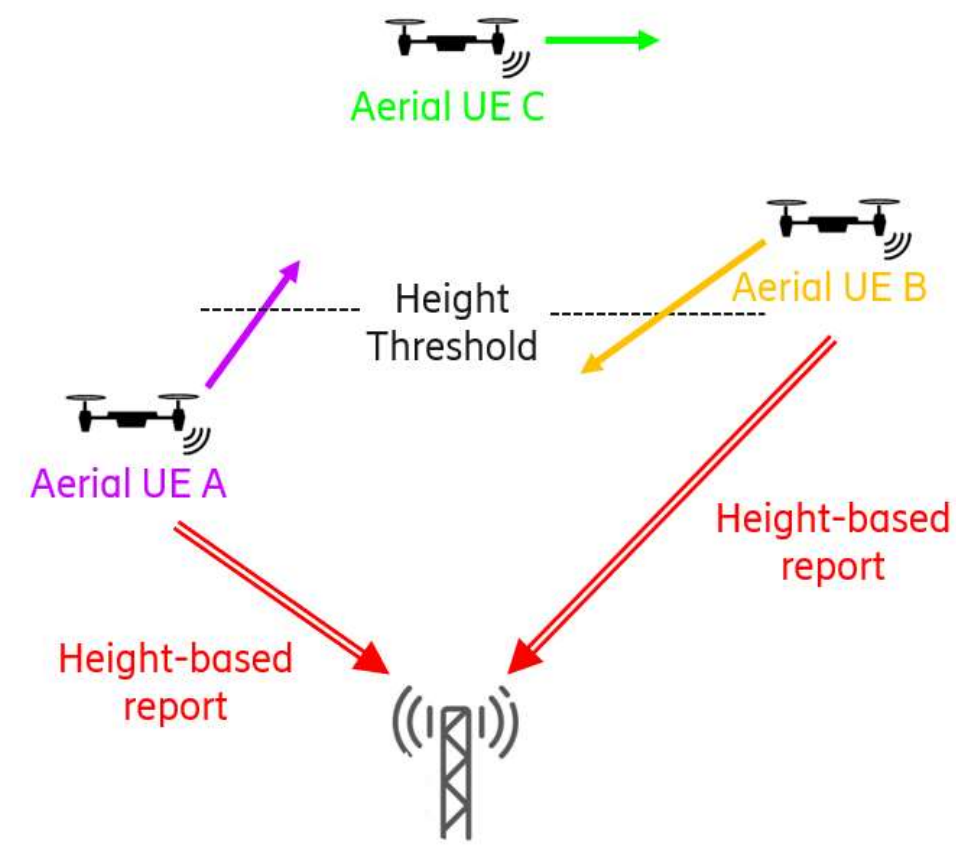

Figure 4. An example illustrating height-based measurement reporting

Figure 4 shows an example of height-based measurement reporting. In this example, all three aerial UEs are configured with the same height threshold. Aerial UE A which is moving upwards reports a height-based measurement when it goes above the height threshold. Aerial UE B which is moving downwards reports a height-based measurement when it goes below the height threshold. Aerial UE C which is moving horizontally does not report a height-based measurement as it does not cross the configured height threshold.

## V. Interference Mitigation

In this section, we describe interference mitigation techniques and the features standardized in the Release-15 work-item phase.

### *A. Uplink Interference Mitigation Solutions Considered During Study-Item Phase*

UE specific fractional pathloss compensation requires the introduction of a UE specific fractional pathloss compensation parameter which is an enhancement to the open loop power control mechanism that existed in LTE up until Release-14 (see

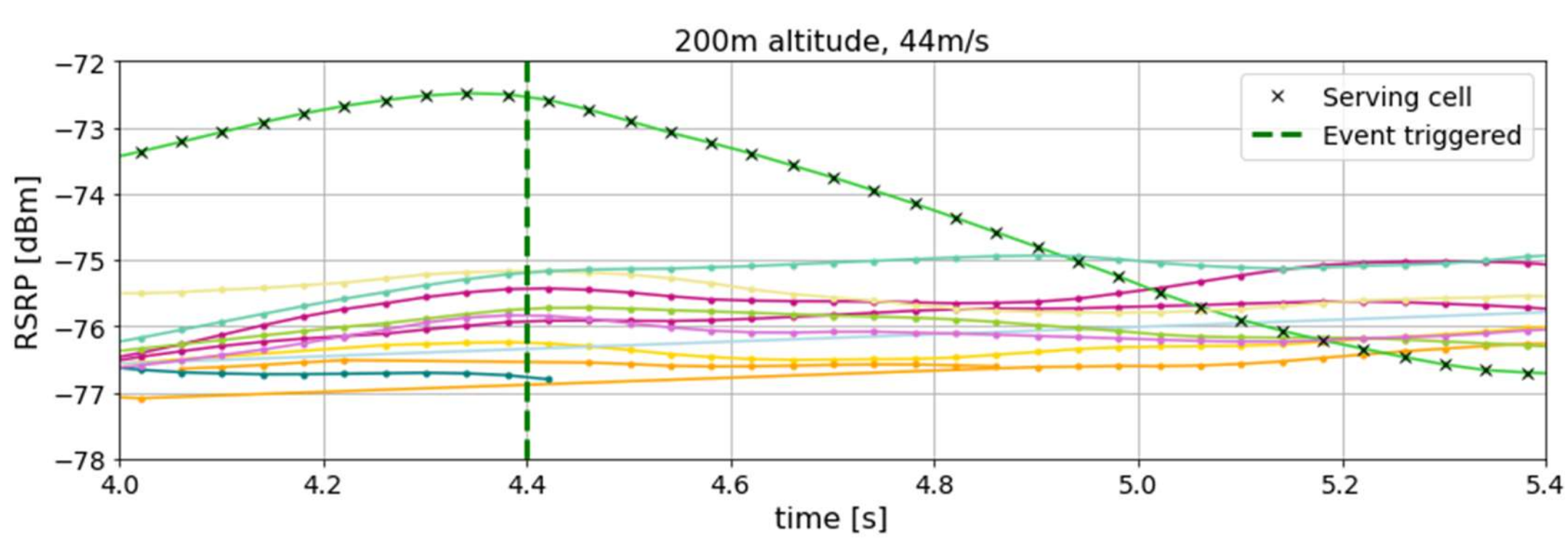

Figure 3. A numerical example for the case where the UE is configured to report a measurement when 4 cells fulfill entry conditions for event A4; the serving cell is indicated by 'x'.

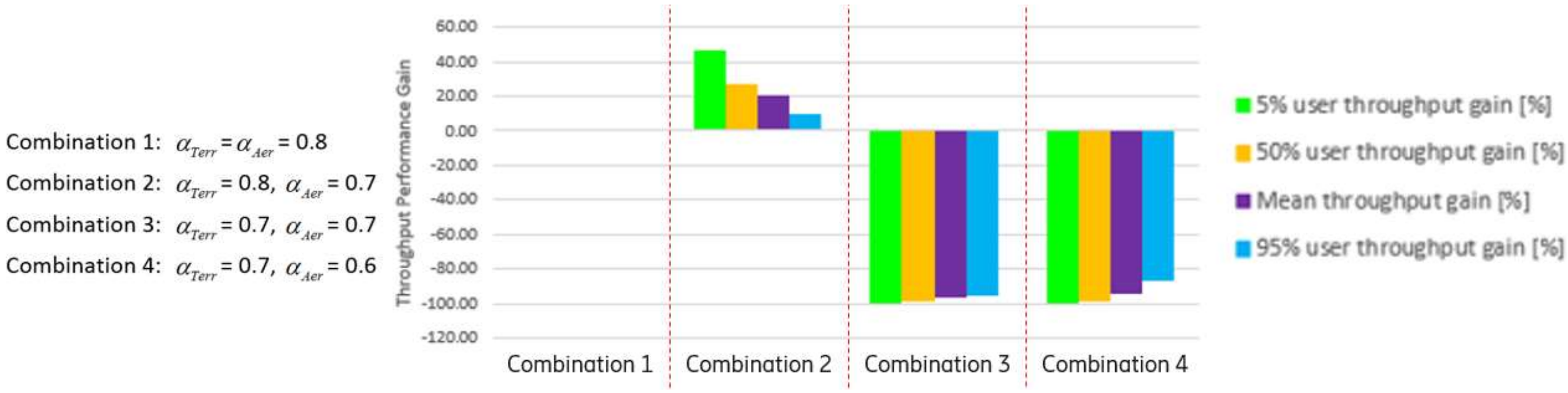

Figure 5. Illustration of uplink throughput performance with the introduction of UE specific fractional pathloss compensation parameter in LTE Release-15.

Clause 5.1 of [14] for details of the existing power control mechanism in LTE). With this technique, the aerial UEs can be configured with a different fractional pathloss compensation factor compared to that configured to the terrestrial UEs. Note that depending on the aerial UE height, different aerial UEs can also be configured with different fractional pathloss compensation factors. A second power control technique is the use of UE specific $P_0$ parameter (note that $P_0$ is an open loop power control parameter specified in LTE and it is the target power level that the eNodeB wants to receive per resource block). With this technique, the aerial UEs can be configured with a different $P_0$ parameter compared to that configured to the terrestrial UEs. Since the $P_0$ parameter can be UE specifically configured in LTE, this technique does not require specification enhancements. However, it was found that the range of values supported in LTE for UE specific $P_0$ may need to be extended.

In closed loop power control based technique, the target received power of the aerial UEs was adjusted considering measurement reports received from both serving and neighbor cells. Closed loop power control may need to cope with potential fast signal change in the sky. Such fast signal changes are possible in the sky given the aerial UEs may be served by the sidelobes of eNodeB antennas. To cope with such fast signal changes, the step size of the transmit power control command may need to be increased which in turn may require specification enhancements to the existing power control mechanism in LTE.

With full dimensional multi-input multi-output (FD-MIMO), multiple receive antennas at the eNodeB can be used to mitigate interference in the uplink. As FD-MIMO is supported in LTE since Release-13, this technique does not require any specification enhancements.

The use of directional antennas at the aerial UE can also help reduce the uplink signal power from the aerial UE in a broad range of angles. This in turn helps reduce the uplink interference caused by the aerial UEs. The following types of aerial UE LOS tracking were considered:

- the antenna direction of the aerial UE is aligned with the direction of travel (DOT) of the UE
- the LOS direction to the serving cell is either ideally tracked by steering its antenna boresight towards the serving cell or non-ideally tracked UE with errors due to practical constraints

As the use of directional antennas at the aerial UE is an implementation issue, this technique does not require any specification enhancements.

### *B. Uplink Interference Mitigation Features Specified during Work-Item Phase*

During the Release-15 work-item phase, a UE specific fractional pathloss compensation parameter was introduced in LTE. With this, the aerial UEs can be configured with a different fractional pathloss compensation factor compared to that configured to the terrestrial UEs. Figure 5 shows the terrestrial UE uplink throughput performance comparison between different combinations of fractional path loss compensation factors configured to the terrestrial and aerial UEs. In Figure 5, the pathloss compensation factors configured to the aerial and terrestrial UEs are denoted by $\alpha_{Terr}$ and $\alpha_{Aer}$, respectively. When compared to combination 1, it can be seen that combination 2 can improve terrestrial UEs' 5-percential uplink throughput by roughly 45% and mean uplink throughput by 20%. In contrast, combinations 3 and 4 result in significant performance degradation. Hence, the results from Figure 5 emphasize that the terrestrial UE performance can be improved significantly with proper configuration of pathloss compensation factors to the aerial and terrestrial UEs.

A second uplink interference mitigation feature specified in the Release-15 work-item phase is the extended range of the UE specific component of the $P_0$ parameter. Up until LTE Release-14, the UE specific component of the $P_0$ parameter had a value range of -8 dB to +7 dB. During the Release-15 work-item phase, the value range of the UE specific component of the $P_0$ parameter is extended to the range of -16 dB to +15 dB to provide better flexibility in setting open loop power control parameters on a UE specific basis.

### *C. Downlink Interference Mitigation*

With FD-MIMO, multiple transmit antenna ports at the eNodeB can be used to mitigate downlink interference to the aerial UEs. The use of directional antennas at the aerial UE can reduce the downlink interference to the aerial UE from a broad range of angles. When the aerial UEs are equipped with more than 2 receive antennas, receive beamforming can be an effective interference mitigation technique in the downlink. In the intra-site JT CoMP (joint transmission coordinated multiple points) scheme, data are jointly transmitted to the UEs from multiple cells that belong to the same site. Coverage extension techniques can enhance synchronization and initial access performance of aerial UEs.

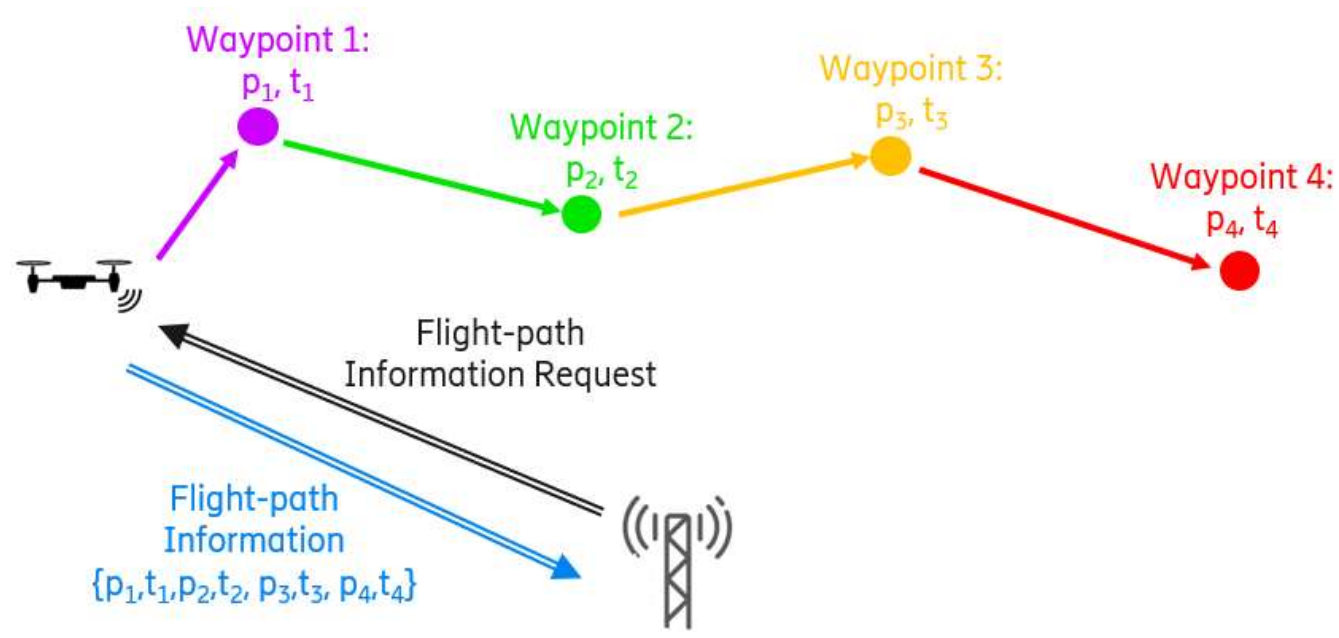

Figure 6. An example illustrating flight path plan reporting

The above interference mitigation solutions studied during Release-15 were either already supported in LTE (FD-MIMO, intra-site JT CoMP, and coverage extension) or implementation-based techniques (directional antennas at the aerial UE and receive beamforming). Hence, it was concluded that these studied solutions were sufficient to guarantee reliable downlink interference mitigation and further downlink interference mitigation features were not introduced during the Release-15 work-item phase.

## VI. MOBILITY AND AERIAL UE IDENTIFICATION

In this section, we discuss mobility performance and identified solutions, followed by a description of the features standardized in the Release-15 work-item phase. We also discuss how to identify aerial UE in a cellular network.

### *A. Mobility Performance and Solutions Identified During Study-Item Phase*

During the Release-15 study, mobility simulations were performed and measurements from field trials were collected [1]. From these results, the mobility performance of an aerial UE is shown to be worse when compared to that of a terrestrial UE especially when the number of aerial UEs is large. Due to the increased downlink interference, the downlink signal to interference plus noise ratio (SINR) for the aerial UEs is much worse than the downlink SINR for the terrestrial UEs. Hence, the aerial UEs may experience more handover failures, more radio link failures, longer handover interruption time, etc. The mobility simulation results showed a better mobility performance for aerial UEs in the RMa-AV scenario than in the UMa-AV scenario. It should be noted however that interference mitigation techniques listed in Section V were not considered in the mobility simulations of the Release-15 study and the use of such techniques is expected to improve the aerial UEs' mobility performance.

Some techniques for improving mobility performance of aerial UEs include (1) enhancements to handover procedure such as conditional handover and handover related parameters considering such as location information, airborne status, flight path plans, etc., and (2) enhancements to existing measurement reporting mechanisms such as new events, enhancements of triggering conditions, etc.

### *B. Flight path Signaling*

During the work-item phase of Release-15, a flight path reporting feature was specified. In this feature, the network can request flight path information from the aerial UE at any time and UE signals the flight path information if it is available at the UE. In order to limit unnecessary signaling, an aerial UE can indicate to the network if it has flight path information available (for example when radio resource control connection is initiated). Upon receiving the flight path information request, the aerial UE reports the flight path information to the network. Figure 6 shows an example of flight path information reporting. As shown in the figure, the flight path information includes the waypoints ($p_1$, $p_2$, ...) and the corresponding time-stamps ($t_1$, $t_2$, ...). The flight path plan can be used by the network to know, for instance, how many drones are to be served in an area. This may help network to plan the used resources.

### *C. Aerial UE Identification*

Depending on country-specific regulations, an aerial UE may need to be identified by the network in order to allow the use of LTE networks for aerial UE connectivity. Another aspect is that there may be drone specific service or charging by the operator.

Aerial UE identification solution can be a combination of user based identification via subscription information and device functionality based identification via LTE radio capability signaling. The mobility management entity (MME) can signal the subscription information to the eNodeB which can include information on whether the user is authorized to operate for aerial usage. In addition, an aerial UE as LTE device can indicate its support of aerial related functions via radio capability signaling to the eNodeB. The combination of the subscription information and the radio capability indication from the UE can be used by the eNodeB to identify an aerial UE, and then perform the necessary control and the relevant functions.

## VII. CONCLUSIONS

In Release-15, 3GPP has dedicated a significant effort during its study on LTE connected drones and concluded that it is feasible to use existing LTE networks to provide connectivity to low altitude drones despite some challenges, as overviewed in this article. The article also sheds light on the features that were specified during the Release-15 work-item phase. The features were specified to enhance interference detection, uplink interference mitigation, mobility performance, and aerial UE identification. Providing efficient and effective connectivity to the aerial UEs while minimizing the impact on terrestrial devices requires a rethinking of many of the assumptions, models, and techniques used to date for cellular system. This article has particularly focused on the 3GPP state-of-the-art findings and solutions on LTE connected drones, although most of the lessons herein would likely apply to any cellular systems (such as 5G) providing connectivity to the sky.